\documentstyle[preprint,aps]{revtex}

\input{psfig}
\begin{document}
\title{A New Method for Computing Topological Pressure}
\author{P\'eter Pollner}
\address{E\"otv\"os University Budapest, \\
Department of Solid State Physics\\
H-1088 Budapest, M\'uzeum krt. 6-8., Hungary\\
e-mail:pollnerp@ludens.elte.hu}
\author{G\'abor Vattay\cite{LAbs}} 
\address{Division de Physique Th\'eorique, Institut de Physique Nucl\'eaire,\\
F-91406 Orsay Cedex, France\\
e-mail:vattay@ipncls.in2p3.fr}
\date{\today}

\maketitle

\mediumtext
\begin{abstract}
The topological pressure introduced by Ruelle and similar quantities
describe dynamical multifractal properties of dynamical systems.
These are important characteristics of mesoscopic systems in the
classical regime. Original definition of these quantities are based on 
the symbolic description of the dynamics. It is hard or impossible to find 
symbolic description and generating partition to a general dynamical system, 
therefore these quantities are often not accessible for further studies. 
Here we present a new method by which the symbolic description can be
omitted. We apply the method for a mixing and an intermittent 
system. 
\end{abstract}
\narrowtext

In recent years the application of the thermodynamic formalism\cite{ruelle} 
(TF) in analyzing dynamical multifractal properties
\cite{Grassberger,Procaccia,Eckman} has been widely accepted. Besides giving an 
illuminating analogy 
with the statistical mechanics, it provides a deeper understanding of 
nonanalytic behavior in the scaling properties of trajectories in dynamical 
systems which can be interpreted as phase transitions\cite{Phase}. 
The topological pressure (TP) plays a central role in the TF. 
The computation of the TP is difficult in general, since the knowledge
of the {\em symbolic dynamics}\cite{Simbol} of the system and its generating 
partition is inevitable. There is no general theory at present, which can 
provide a systematic and numerically realizable method to construct a 
generating partition to a general dynamical system\cite{Bene}. Therefore the TP 
has
been computed mostly for low dimensional maps and billiard systems,
where the symbolic dynamics is accessible,
or has been computed with averaging the generalized Lyapunov 
exponents\cite{Gaspard95}.
 In {\em mesoscopic}
devices like dots, antidots and wells Hamiltonian dynamical systems with 
smooth potentials play significant role. It would be very enlighting
to apply TF for such non-trivial systems.  
In this Letter we 
introduce a new technique based on a suitably defined correlation
function\cite{CV}
to measure the TP. The advantage of the method is
that the detailed knowledge of the system is not required
and the calculation can be made with low computational demand for either hyperbolic
or mixing systems.
 The new method makes possible to measure TP by following one single trajectory 
of an ergodic system.

The TP for maps can be defined as the logarithm 
$P(q)=log z_0(q)$ of the leading zero
$z_0(q)$ of the Ruelle zeta function
\begin{math}
1/\zeta(z,q)=\prod_p\left(1-\frac{z^{n_p}}{|\Lambda_p|^q}\right),
\label{def}
\end{math}
where the infinite product goes for the non-repeating or primitive periodic
orbits of the system, $\Lambda_p$ is the largest eigenvalue of 
its stability matrix ${\bf J_p}$ and $n_p$ is its period. 
This definition has been extended for continuous flows in Refs.\cite{ruelle,CE}. 
The TP is the leading zero $P(q)=s_0(q)$ of the zeta function
\begin{math}
1/\zeta(s,q)=\prod_p\left(1-\frac{e^{sT_p}}{|\Lambda_p|^q}\right),
\end{math}
where $T_p$ is the time period of the periodic orbit. The TP is
equivalent with the free energy\cite{Bohr} $q F(q)$ and also with the
R\'enyi entropies\cite{Procaccia} in hyperbolic systems which are defined 
in a slightly different way in the literature. 
In the definitions above, the problem to find the symbolic dynamics is hidden,
since it is necessary to locate the periodic orbits in the system.
At certain values of $q$ the TP has a special meaning\cite{transient}. For $q=0$
the quantity $-P(0)$ is the 
topological entropy per unit time. For $q=1$ it is zero
for bound systems, it is the escape rate for open systems and its derivative
gives the metric entropy per unit time. The parameter $q$ where the TP 
is  zero ($P(q)=0$) yields the fractional part of the fractal dimension of 
the repeller in scattering systems. The time 
$-1/P(1/2)$ gives a lower
bound for the quantum resonance lifetimes in semiclassical approximation
for open systems\cite{CRR93,gasp,GA}. 

Zeta functions can be related to transfer operators.
In Ref.\cite{CV} we have introduced a transfer operator in arbitrary finite 
dimensions, whose largest eigenvalue is related to the TP. Here we briefly
repeat the main idea. 

The dynamical system has to be extended to
the tangent space of the flow,
where stability of the orbits is multiplicative.
We adjoin
the $d$-dimensional transverse tangent space
${\bf \xi}\in TU_x$, ${\bf \xi}(x) \cdot {\bf v}(x)=0$,
to the ($d$+1)-dimensional dynamical evolution
space $x\in U\subset {\bf R}^{d+1}$.
The dynamics in the $(x,\xi) \in U \times TU_x$ space
is governed by the system of
equations of variations\cite{arnold73}:
\[
\dot{x}={\bf v}(x) \,, \quad
\dot{{\bf \xi}}={\bf Dv}(x){\bf \xi }\, .
\]
Here ${\bf Dv}(x)$ is the transverse derivative matrix of the flow.
We write the solution as
\begin{equation}
x(t)=f^t(x_0) \,, \quad
{\bf \xi}(t)={\bf J}^t(x_0) \cdot {\bf \xi}_0 \, ,
\label{xit}
\end{equation}
with the tangent space vector ${\bf \xi}$ transported by
the transverse stability matrix\ ${\bf J}^t(x_0) = \partial x(t)/ \partial
x_0$.
In order to determine the length of the vector ${\bf \xi}$
we introduce a {\em signed norm}.
An example is the function
\begin{equation}
g: TU_x\rightarrow {\bf R},\ \ \ g \left( \begin{array}{c}
        \xi_1 \\
        \xi_2 \\
        \cdots \\
        \xi_d
\end{array} \right)= \xi_d \,.
\label{proj_norm}
\end{equation}
Any vector ${\bf \xi}\in TU_x$ can now be represented by the product
$
{\bf \xi}=\Lambda {\bf u}
$,
where $ {\bf u}$ is a unit vector in the
 signed norm: $g({\bf u})=1$, and the factor
\begin{equation}
\Lambda^t(x_0,{\bf u}_0)=g(\xi(t)) = g({\bf J}^t(x_0) \cdot {\bf u}_0)
\label{lamb_def}
\end{equation}
is the multiplicative ``stretching'' factor
\[
\Lambda^{t'+t}(x_0,{\bf u}_0)=\Lambda^{t'}(x(t),{\bf u}(t))
                                \, \Lambda^t(x_0,{\bf u}_0).
\]
The ${\bf u}$ evolution constrained to $ET_{g,x}$, the space of unit
tangent vectors transverse to the flow ${\bf v}$,
is given by rescaling of (\ref{xit}):
\begin{equation}
{\bf u}'
=R^t(x,{\bf u})=
\frac{1}{\Lambda^t(x,{\bf u})} {\bf J}^t(x) \cdot {\bf u} \, .
\label{R}
\end{equation}
Eqs. (\ref{xit}), (\ref{lamb_def}) and (\ref{R})
enable us to define a {\em multiplicative} evolution operator on
the extended space $ U \times ET_{g,x}$
\begin{equation}
{\cal L}^t(x',{\bf u}';x,{\bf u})=
\delta(x'-f^t(x))
{\delta({\bf u}'-R^t(x,{\bf u})) \over |\Lambda^t(x,{\bf u})|^{q-1} }
\, .
\label{tran_op}
\end{equation}
This operator is the generalization of the well-known transfer operator
$\hat{L}_{q}(x,y)=\mid f'(y) \mid^{1-q}\delta(x-f(y)),$
 for arbitrary dimensions
and continuous time. In analogy with the one dimensional problem, 
we would like to determine its leading eigenvalue in order to get the TP.

Our new method uses  the relation of  the generalized transfer operators
introduced above and the correlation decay in chaotic systems. Let's take two 
arbitrary smooth 
observables $A(x,{\bf u})$ and $B(x,{\bf u})$ depending on the real variables 
$x$ and on  the tangent space variables ${\bf u}$. We can define their
generalized correlation function as  
\begin{equation}
C_{AB}^{q}(t)=\langle A(x,{\bf u})\mid \Lambda^t(x,{\bf u})\mid^{1-q}B(f^t(x),
R^t(x,{\bf u}))\rangle,
\end{equation}
where the average is uniform in the extended space
\begin{equation}
\langle A(x,{\bf u})\rangle=\int d{\bf u}dxA(x,{\bf u}) \label{aver}.
\end{equation} 
With the help of the operator (\ref{tran_op}), the correlation function can 
be rewritten as
\begin{equation}
C_{AB}^{q}(t)=\langle A(x,{\bf u})
\int d{\bf u}'dx'{\cal L}^t(x,{\bf u};x',{\bf u}')B(x',{\bf u}')\rangle .
\end{equation} 
We can introduce the eigenvalues $-s_n(q)$ and eigenfunctions (which are
distributions in general)
$\psi_n(x,{\bf u})$
of the stationary problem of the operator (\ref{tran_op}) 
\begin{equation}
e^{-s_i(q) t}\psi_n(x,{\bf u})=\int d{\bf u}'dx'{\cal L}^t(x,{\bf u};x',{\bf 
u}')
\psi_n(x',{\bf u}').
\end{equation}
 
For {\em Axiom A systems}\cite{Rugh92} one can expand $B(x',{\bf u}')$ on the 
eigenbasis:
\begin{equation}
B(x',{\bf u}')=\sum_n b_n \psi_n(x',{\bf u}'). \label{deko}
\end{equation}
The correlation function then becomes
\begin{equation}
C_{AB}^{q}(t)=\sum_n e^{-s_i(q) t} b_n
\langle A(x,{\bf u})\psi_n(x,{\bf u})\rangle .
\end{equation}
For large $t$ this sum is dominated by the term corresponding to the
largest eigenvalue $-s_0(q)$
\begin{equation}
C_{AB}^{q}(t)\sim  e^{-s_0(q) t} C,\label{expo}
\end{equation}
where the constant is
$C=b_0\langle A(x,{\bf u})\psi_0(x,{\bf u}))\rangle$ .
For general, non-Axiom A systems, the decomposition  (\ref{deko})
is not valid, since the eigenfunctions (distributions) not necessarily
span the full function space. Nevertheless, at long times, the decay is
always dominated by the largest eigenvalue in accordance with (\ref{expo}).

Therefore the TP can be read off from the semi-logarithmic plot
of the correlation function for sufficiently large time $t$.

For an ergodic system the phase space average (\ref{aver}) can be 
computed as a suitable time average for a sufficiently long ergodic
trajectory. If, for the simplicity, we choose the special observables 
$A(x,{\bf u})=1$ and $B(x,{\bf u})=\varrho(x)$, where $\varrho(x)$ is 
the equilibrium probability distribution of the system, the correlation 
function becomes
\begin{equation}
C_{1\varrho}^{q}(t)=\lim_{T\rightarrow\infty}\frac{1}{T}\int_0^T
dt' \mid \Lambda^{t'+t}(x_0,{\bf u}_0) \mid^{1-q}\label{mert}
\end{equation}  
where $x_0$ and ${\bf u}_0$ represent an initial condition for a 
long ergodic trajectory. 

To implement the technique for a two dimensional classical
problem, 
we need a convenient choice of the $g(\xi)$ function. For 
2-dimensional Hamiltonian dynamics
let the  2-dimensional Poincar\'e section return map be $x_{i+1}=f(x_{i})$.
The stability matrix\ of cycle $p$ is a product of the
$2\times 2$ stability matrices
\[ {\bf J}_j= \left( \begin{array}{cc}
                      A_j & B_j \\
                      C_j & D_j
\end{array} \right)
\, ,
\]
where $A_j = \partial f_1(x_j) / \partial x_1$, and so on.
Assume the signed norm (\ref{proj_norm}) and multiply an
initial unit vector
by the first stability matrix\ in the product. The resulting vector can be
written as
\[ {\bf J}_1 \left( \begin{array}{c} \kappa_1 \\ 1 \end{array} \right)
= (C_1\kappa_1+D_1)
\left( \begin{array}{c}
        \frac{A_1\kappa_1+B_1}{C_1\kappa_1+D_1} \\
         1 \end{array} \right)
\, .
\]
Hence the dynamics acts on the unit vectors as a
rational fraction transformation
\begin{equation}
\kappa_{k+1}=R(x_k,\kappa_k)=
\frac{A_k\kappa_k+B_k}{C_k\kappa_k+D_k},
\label{racfrac}
\end{equation}
with the signed norm (\ref{lamb_def}) of the iterated vector given by
\[\Lambda^{n}(x_{1},\kappa_{1}) =
\prod_{i=1}^{n}(C_i\kappa_i+D_i)
\, .
\]
In the case of 2-dimensional billiards,
$\kappa_n$ is the Bunimovich-Sinai curvature\cite{BS80}.
The Bunimovich-Sinai curvature $\kappa(t)$ is the local curvature of the front
(horocycle) formed around a central trajectory by nearby orbits 
started from a common point of origin with the same energy. 
For a periodic orbit $\kappa_{n_p}=\kappa_0$,
the unit vector is an eigenvector of
the stability matrix, and the corresponding eigenvalue is
$\Lambda_p=\prod 
 (C_i\kappa_i+D_i)$.
The $\kappa$ can be defined also
for continuous time\cite{japan}. Instead of the rational fractional 
transformation (\ref{racfrac}), we get a differential equation for $\kappa(t)$.
The "stretching factor" becomes the integral 
\begin{equation}
\Lambda^t(x_0,\kappa_0)=\exp\left(\int_0^t \kappa(\kappa_0,x_0,t')dt'\right).
\end{equation}
The time evolution of $\kappa(t)$  
for 2-dimensional Hamiltonian systems with hamiltonian 
$H=\frac{1}{2}(p_x^2+p_y^2)+U(x,y)$ can be 
derived easily: An infinitesimal configuration space volume $V_0$, formed around 
a central 
trajectory by nearby orbits started from a common point of origin with the same 
energy,
is stretched after time $t$  with the factor $\Lambda^t(x_0,\kappa_0)$ orthogonal 
to the central trajectory and with $\sqrt{\frac{2(E-U(x(t))}{2(E-U(x(0))}}$ 
along the trajectory, due to the 
change of the  velocity along the trajectory. The change of the volume in unit 
time can be written as $\frac{dV_t}{dt}=\lambda(t)V_t$, where the expansion rate is 
the sum of the orthogonal and the parallel stretching rates: 
$$\lambda(t)=\kappa(t)+\frac{d}{dt}\frac{1}{2}\log(2(E-U(x(t)))).$$ 
Computation of
$\kappa(t)$ is then reduced to the problem of finding the expansion rate of 
volumes.
This can be recovered by investigating the evolution of
trajectories, which deviate infinitesimally $\delta x(t)$ from the central 
trajectory
$x(t)$. These are described by the  linearized Newton equation 
$\delta\ddot{x}(t)=
-{\bf D^2}U(x(t))\delta x(t)$, where ${\bf D^2}U(x(t))$ is the second
derivative matrix of the potential. Since the trajectories are restricted to the 
same energy surface as the central trajectory, we get the additional 
constraint:
$\dot{x}(t)\cdot \delta\dot{x}(t)+\nabla U(x(t))\delta x(t)=0$.
The evolution in two dimensions can be reformulated to the linear  form  
\begin{equation}
\delta\dot{x}(t)={\bf M}(t)\delta x(t),\label{M}
\end{equation}
where the four elements of the
two-by-two matrix ${\bf M}(t)$ can be constructed from the four vector 
components of
$\delta\dot{x}(t)$ and $\delta x(t)$. The evolution of the matrix ${\bf 
M}(t)$
can be determined from the linearized Newton equation: 
$\dot{{\bf M}}(t)=-{\bf M}^2(t)-{\bf D^2}U(x(t))$,
and the energy conservation constraint yields:
$\dot{x}(t){\bf M}(t)+\nabla U(x(t))=0$.
The expansion rate of the configuration space volume is the divergence of the 
velocity
field (\ref{M}), which is the trace of the matrix  ${\bf M}(t)$: 
$\lambda(t)=Tr {\bf M}(t)$.

For the $\kappa(t)$ we get:
\begin{equation}
\dot \kappa=-\kappa^2
  -3 \frac{(p_y\partial_x U-p_x\partial_y U)^2}
        {(p_x^2+p_y^2)^2} 
  -\frac{p_y^2 \partial_{xx} U
         -2 p_x p_y \partial_{xy} U
         +p_x^2\partial_{yy} U}
         {p_x^2+p_y^2}
\end{equation}

The numerical calculations have confirmed the applicability
of the method. For concrete calculations we have chosen the Hamiltonian of the 
Anisotropic Kepler Problem (AKP)\cite{Gutzwiller,tanner} 
$$H=\frac{1}{2}(p_x^2+p^2_y)-\frac{1}{(x^2+\alpha y^2)^{1/2}},$$
which describes a Bloch electron with anisotropic mass tensor
in a Coulomb potential and the Hamiltonian
$$H=\frac{1}{2}(p_x^2+p^2_y)+\frac{1}{2}x^2y^2$$ which 
can be considered as a model of a soft wall mesoscopic cross 
junction\cite{junk}.
We have measured the
correlation function (\ref{mert}) for the AKP at parameter
$\alpha=5.0$ and for the $\frac{1}{2}x^2y^2$ potential at energy $E=-0.5$ up to
$T=13000$ and $E=+0.5$ up to $T=30000$ respectively,
for several initial conditions while we integrated the equations with 
$dt=0.001$ using the fourth order Runge-Kutta\cite{numrec} method
with adaptive step-size control. Both systems can be considered practically
ergodic, since regular islands cover less than $0.005$ percent of their phase
space\cite{ergo}. The main difference between these systems is that the 
AKP is a mixing one at the chosen value of $\alpha$, while the other one 
is intermittent.

For the AKP the correlation function (\ref{mert}) reached
the asymptotic behavior very quickly for all $q$ (Fig. 1b),
while for the intermittent system one can see the effect of the 
critical slowing down\cite{kritlass} at $q>1$ (Fig. 2b).
The measured TP functions are depicted on Fig. 1a and 2a. 

We hope, that the method presented here
can be applied for many dynamical systems, where other methods have failed
so far.

The authors are grateful to P. Cvitanovi\'c, H. H. Rugh,
T. T\'el and P. Sz\'epfalusy for discussions. G. V. thanks the European
Community Nonlinear Network the fellowship during his stay in Orsay.
This work was partially supported by OTKA T17493, F17166 and the Foundation for 
the Hungarian Higher Education and Research.

\begin{figure}
\centerline{\strut
\psfig{file=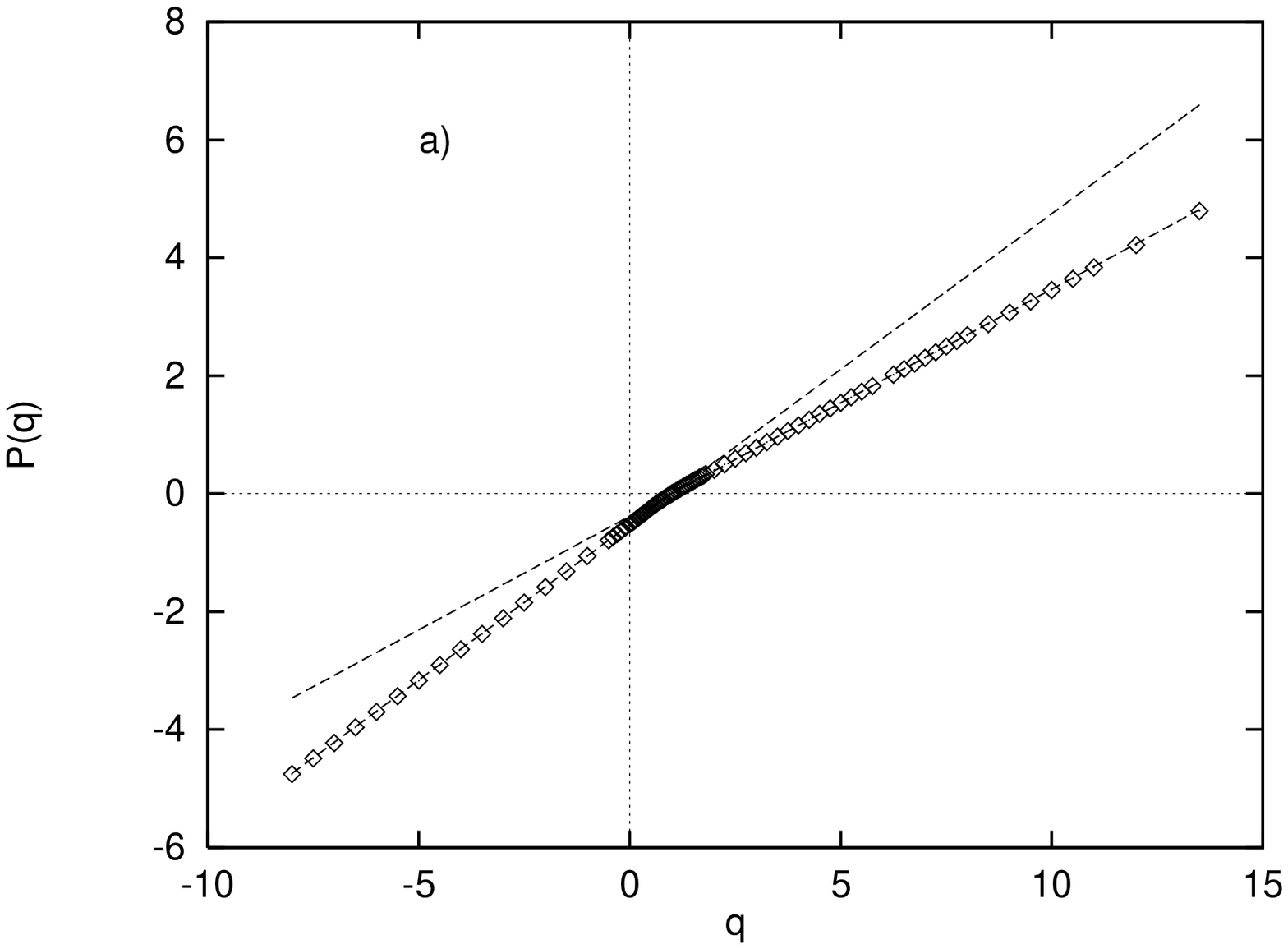,height=65mm}
\psfig{file=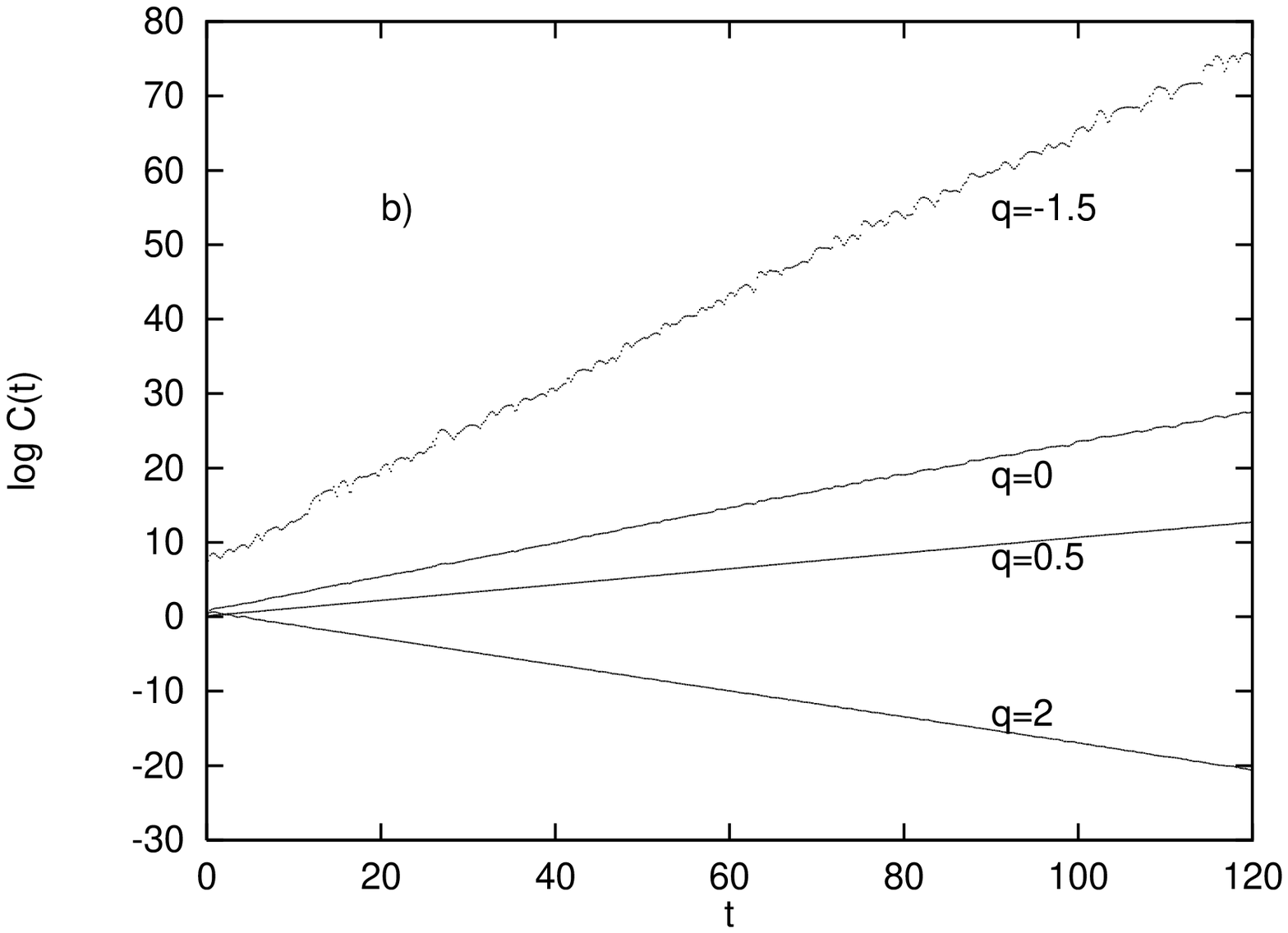,height=65mm} }
\caption{a) The measured Topological Pressure (per unit time) function for the
AKP as a function of $q$ with the asymptotes. b) The common logarithm of the 
measured correlation function of the AKP  
as a function of time $t$.
Different straight lines correspond to different values of $q$, 
due to the fast convergence to the asymptotic behavior.}
\end{figure}

\begin{figure}
\centerline{\strut                                                              
\psfig{file=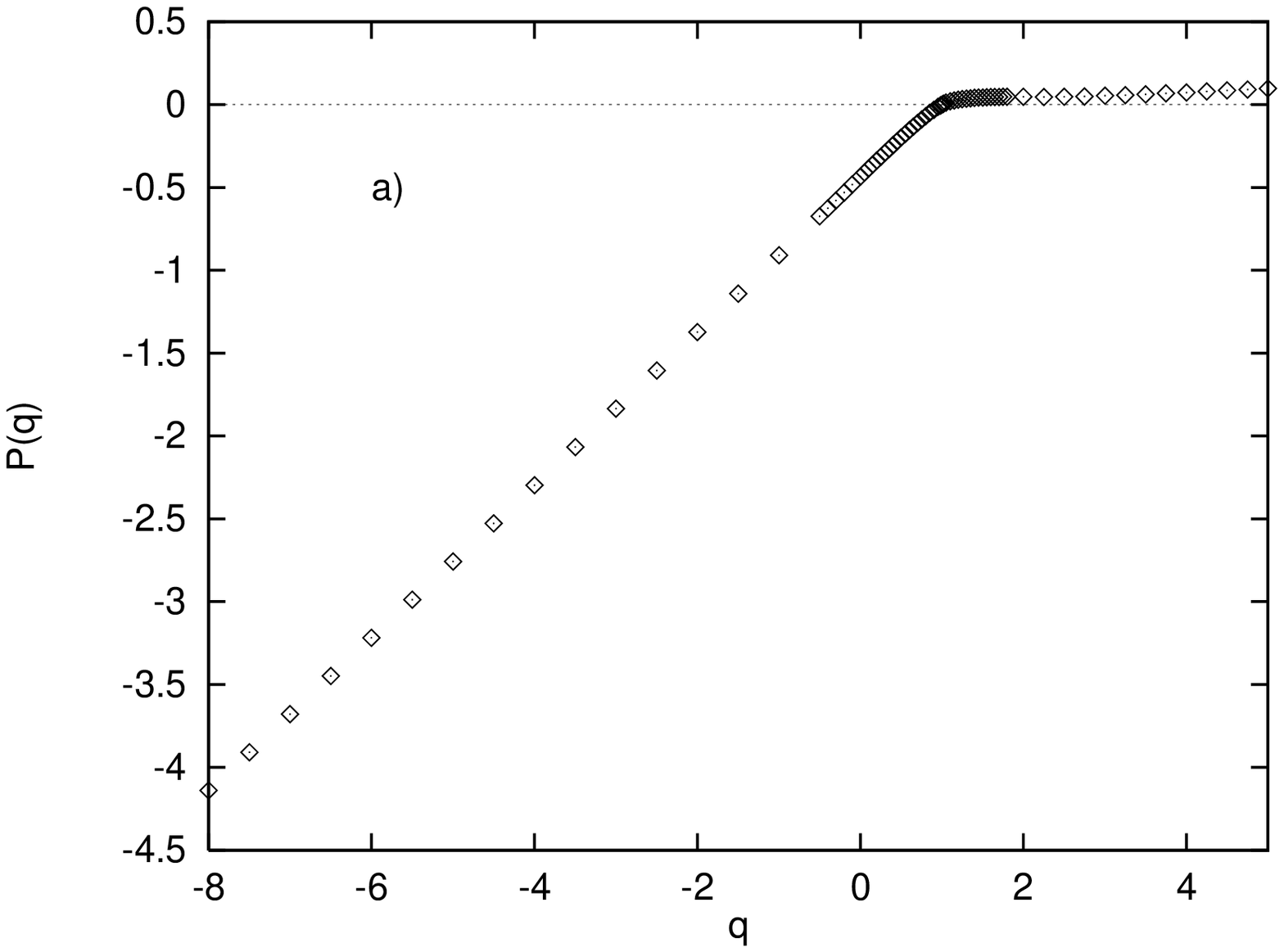,height=65mm}  
\psfig{file=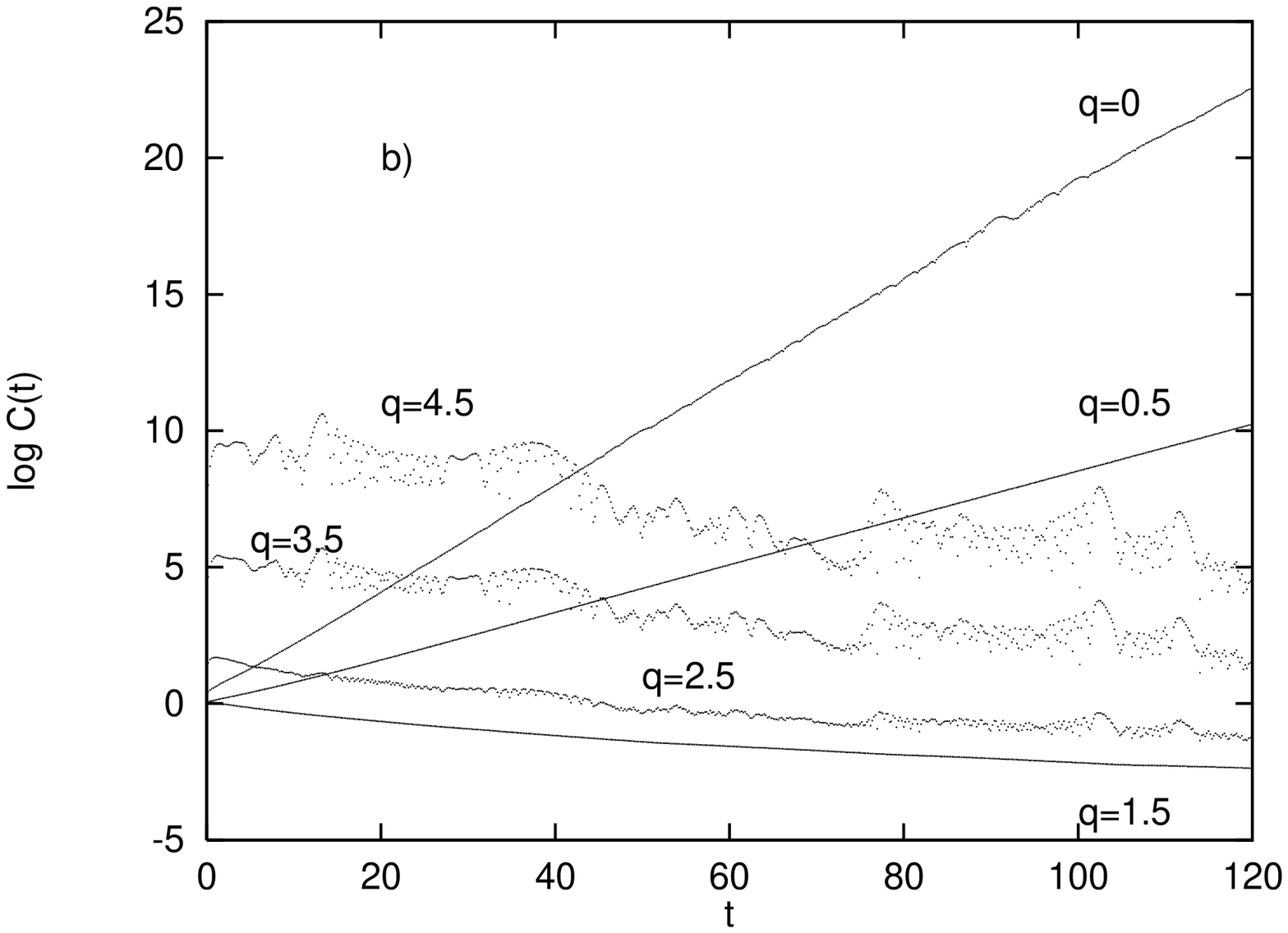,height=65mm} }
\caption{a) The measured Topological Pressure (per unit time) function for the 
$\frac{1}{2}x^2y^2$ potential as a function of $q$. 
b) The common logarithm of the correlation function of 
the $\frac{1}{2}x^2y^2$ potential as a function of time $t$. 
One can observe the critical slowing down of the convergence as 
$q$ becomes greater than 1}
\end{figure}

\begin{references}
\bibitem[*]{LAbs} On leave from E\"otv\"os University Budapest, Department of 
Solid State Physics 
\bibitem{ruelle} Y. G. Sinai, Usp. Mat. Nauk. {\bf 27}, 21 (1972);
R. Bowen, Lect. Notes Math {\bf 470}, 1 (1975);
D. Ruelle, {\em Statistical Mechanics, Thermodynamic
Formalism} (Addison-Wesley, Reading, MA, 1978)
\bibitem{Grassberger} H. Kantz and P. Grassberger, Physica {\bf D17}, 75 (1985)
\bibitem{Procaccia} P. Grassberger and I. Procaccia, Phys. Rev. {\bf A28},
2591 (1983)
\bibitem{Eckman} J. P. Eckmann and I. Procaccia, Phys. Rev. {\bf A34}, 659
(1986)
\bibitem{Phase} P. Cvitanovi\'c, in {\em Group Theoretical Methods in Physics},
ed.: Gilmore (World Scientific, Singapore 1987); P. Sz\'epfalusy and T. T\'el,
Phys. Rev. {\bf A35}, 477 (1987); D. Katzen, I. Procaccia, Phys. Rev. Lett. 
{\bf 58}, 169 (1987)
\bibitem{Simbol}  P. Cvitanovi\'c, G. H. Gunarate and I. Procaccia, 
Phys. Rev. {\bf A38}, 1503 (1988)
\bibitem{Bene} J. Bene, P. Sz\'epfalusy and \'A. F\"ul\"op,  Phys. Rev. 
{\bf A40}, 6719 (1989)
\bibitem{Gaspard95} P. Gaspard and F. Baras, Phys. Rev. E {\bf 51}, 5332 (1995)
\bibitem{CV} P. Cvitanovi\'c and G. Vattay,
Phys. Rev. Lett. {\bf 71}, 4138 (1993)
\bibitem{CE} Cvitanovi\'c and Eckhardt, J. Phys. {\bf A24}, L237 (1991)
\bibitem{Bohr} T. Bohr and D. Rand, Physica {\bf D25}, 387 (1987)
\bibitem{transient} T. T\'el, {\em Transient chaos}, {\em in Directions in
Chaos}, Vol.3, 149 , ed.: Bai-lin Hao,   (World Scientific, Singapore, 1990)


\bibitem{CRR93} P. Cvitanovi\'c, P.E.~Rosenqvist, H.H. Rugh, and G. Vattay,
CHAOS {\bf 3} (4), 619 (1993).



\bibitem{gasp} P. Gaspard and S.A. Rice,
{\em J. Chem. Phys. \bf 90\rm, 2225 (1989); \bf 90\rm, 2242 (1989);
 \bf 90\rm, 2255 (1989).}


\bibitem{GA} P. Gaspard and D. Alonso Ramirez,
        {\em Phys. Rev. \bf A 45},  8383 (1992).

\bibitem{BGS80} G. Bennettin, L. Galgani and  J.-M. Strelcyn,
        {\em Meccanica \bf 15}, 9 (1980).

\bibitem{arnold73} V.I.~Arnold, {\em Ordinary Differential
        Equations} (MIT Press, Cambridge, Mass. 1978)

 \bibitem{Rugh92} H.H. Rugh,
        {\em Nonlinearity \bf 5}, 1237 (1992). See also ref.~\cite{CRR93} for
        a discussion of the applicability of the theorem.
\bibitem{Gutzwiller} M. C. Gutzwiller, J. Math. Phys. {\bf 12},
343 (1971); {\em Chaos in Classical
and Quantum Mechanics} (Springer-Verlag, New York, 1990)
\bibitem{tanner}G. Tanner, D. Wintgen, Chaos {\bf 2}, 53, (1992);
G. Tanner, D. Wintgen, Chaos {\bf 5}, 1325 (1993)
\bibitem{BS80} L. Bunimovich and Ya.G. Sinai,
        {\em Comm. Math. Phys. \bf 78}, 247 (1980);
        {\bf 78}, 479 (1980);
        {\em Erratum, ibid. \bf 107}, 357 (1986).
\bibitem{japan} B. Eckhardt and D. Wintgen, J. Phys. A 24, 4335 (1991);

\bibitem{junk} 
T. Geisel, R. Ketzmerick and O. Schedletzky,
Phys. Rev. Lett. {\bf 69}, 1680 (1992)
\bibitem{numrec} {\em Numerical Recipes} ed.: W.\ H.\ Press, S.\ A.\ Teukolsky,
W.\ T.\ Vetterling, B.\ P.\ Flannery (Cambridge Univ., 1986)

\bibitem{ergo} P. Dahlqvist, G. Russberg, Phys. Rev. Lett. {\bf 65},
2837 (1990)



\bibitem{kritlass} A. Csord\'as, P. Sz\'epfalusy,
                {\em Phys. Rev.} {\bf A} 38, 2582 (1988)
\end{references}
\end{document}